\documentclass[showpacs,twocolumn,amsmath,amssymb]{revtex4}
\usepackage[T1]{fontenc}
\usepackage[latin1]{inputenc}
\usepackage[dvips]{graphicx}
\usepackage{color}

\bibpunct{}{}{,}{s}{}{$^,$}%

\renewcommand{\vec}[1]{\boldsymbol{#1}}
\newcommand{\grad}{\vec{\nabla}}
\newcommand{\curl}{\vec{\nabla}\times}
\renewcommand{\div}{\vec{\nabla}\cdot}

\newcommand{\f}[2]{\frac{#1}{#2}}

\newcommand{\rey}{\mathrm{Re}}

\newcommand{\eps}{\varepsilon}

\newcommand{\difft}[1]{\partial_t #1}

\newcommand{\ddiffyy}[1]{\partial^2_y #1}

\begin{document}

\title{On the existence of two-dimensional nonlinear
   steady states\\ in plane Couette flow}

\author{F. Rincon}
\email{F.Rincon@damtp.cam.ac.uk}
\affiliation{Department of Applied Mathematics and Theoretical Physics,
University of Cambridge,\\ Centre for Mathematical Sciences, Wilberforce
Road, Cambridge CB3 0WA, United Kingdom}%

\date{\today}

\begin{abstract}
The problem of two-dimensional steady nonlinear dynamics in plane
Couette flow is revisited using homotopy from either plane
Poiseuille flow or from plane Couette flow perturbed by a small
symmetry-preserving identity operator. Our results show 
that it is not possible to obtain the nonlinear plane Couette flow
solutions reported by Cherhabili and Ehrenstein [Eur. J. Mech. B/Fluids, 
14, 667 (1995)] using their Poiseuille-Couette homotopy. We also demonstrate 
that the steady solutions obtained by Mehta and Healey [Phys. Fluids,
17, 4108 (2005)] for 
small symmetry-preserving perturbations are influenced by an artefact of the
modified system of equations used in their paper. However, using a
modified version of their model does not help to find plane Couette
flow solution in the limit of vanishing symmetry-preserving
perturbations either. The issue of the existence of two-dimensional
nonlinear steady states in plane Couette flow remains unsettled.
\end{abstract}


\maketitle
  
\section{Introduction}
Even though plane Couette flow is known to be linearly stable for all values of the
Reynolds number $\rey$, transition to turbulence is observed
experimentally and numerically at quite modest values of $\rey\simeq
350$ in this flow \citep{bech95}. As a consequence, the quest for
nonlinear instabilities has been the subject of many
studies. Since there is no direct connection between the basic state
and potential nonlinear branches of solutions, homotopy (continuation)
from other flows subject to linear instabilities has been a very popular method to
discover such branches.  Exact three-dimensional solutions have been reported
by \citet{nagata90} using homotopy from linearly unstable rotating plane
Couette flow, by \citet{clever92} using homotopy from a convectively
unstable plane Couette flow and by \citet{waleffe01,waleffe03} using a
forcing homotopy.
% : by imposing
% initially an artificial body force to maintain streamwise rolls and streaks
% structures, \citet{waleffe01} was able to trigger a three-dimensional
% inflectional instability which takes on the role of the original force
% and to sustain the rolls, thus leading to a self-sustaining process in the
% limit of zero forcing. 
% A similar procedure had been used previously by
% \citet{nagata88}, \citet{nagata90}, who had found the same three-dimensional
% nonlinear standing waves solutions for non-rotating plane Couette flow
% using  nonlinear continuation from three-dimensional nonlinear
% solutions obtained for a linearly unstable slowly
% spanwise-rotating plane Couette flow  (wavy Taylor vortices). 
These three-dimensional states  match very well experimental
data on transition to turbulence in plane Couette flow, but other
solutions, including two-dimensional ones, may well be participating in
phase-space dynamics and in the transition process. 

Another kind of homotopy was suggested by \citet{milinazzo85}
to discover  two-dimensional nonlinear solutions in plane
Couette flow by connecting the nonbifurcating no-slip plane Couette
flow to the no-slip plane Poiseuille flow,  which becomes linearly unstable at
$\rey=5772$ (\citet{zahn74,herbert76}). Starting from the
travelling-wave solutions in this flow and progressively changing the
original background flow from a Poiseuille to a Couette profile,
\citet{milinazzo85} attempted to continue these
solutions to a pure Couette flow, but they failed to achieve this
goal. Their lack of success may have to do with resolution:
\citet{ehren95} (hereinafter CE95) used the same approach with more
Fourier modes in the streamwise direction and claimed to have identified
steady two-dimensional localized nonlinear solutions for the Couette
flow. \citet{ehren97} subsequently computed the stability of these
two-dimensional steady states and reported the discovery of
three-dimensional nonlinear  structures which are different from those
found by \citet{nagata90}.

Recently, \citet{mehta05} (hereinafter MH05) suggested yet another
interesting homotopy to compute two-dimensional nonlinear states in
plane Couette flow. They
argued that the homotopy from Poiseuille to Couette flow was not
symmetry-preserving, so that possible nonlinear states relying on the
$O(2)$ symmetry of the Couette flow might not be found by continuation from
flows that do not have this symmetry. Instead, they proposed a
symmetry-preserving  homotopy based on small perturbations of
the linearization of the Couette operator that causes
the least stable eigenvalues of the plane Couette flow
%, which are known
%to be near-degenerate (\textit{i.e.} they can be made as close as
%desired to the imaginary axis by increasing $\rey$), 
to become unstable.
Doing so, they hoped to be able to track the nonlinear branch
associated with this bifurcation point down to zero perturbation and to
obtain two-dimensional nonlinear states for the unperturbed plane Couette flow.
% One good reason to believe that this approach may be successful is that
% the spectra of non-normal operators such as the one obtained from
% the Couette linearization are known to be very sensitive to small
% perturbations\citep{trefethen93}, which suggests a very rich
%underlying dynamical behaviour.  
For small perturbation parameters, they discovered steady solutions
looking similar to the solutions of CE95. MH05 notably describe
the numerical appearance of so-called ``frozen-waves'' with phase
speed $c=0$ in perturbed Couette flow starting from bifurcating travelling-wave
solutions at $\rey>390$. CE95 had also noticed the numerical appearance
of solutions with $c=0$ in their calculations. However, these solutions 
cannot be followed down to zero perturbation (exact plane Couette
flow), which raises the questions of the existence of the solutions
reported by CE95 and of the connections between their findings and those
of MH05.

Considering that the case is quite unclear, we attempt to address some
of these issues in this study. We first revisit the Poiseuille-Couette
homotopy problem and argue that the solutions obtained by CE95 are
spurious. Then, we clarify the nature of the steady solutions found by
MH05 at large $\rey$ for small values of their perturbation parameter
$\eps$. We show that a direct connection exists between travelling and
steady solutions for the perturbed flow and demonstrate that their high
Reynolds number steady solutions are influenced by an artifact in the
model that introduces a mean flow eigenmode. We obtain solutions for
small perturbations using a revised symmetry-preserving homotopy and
show that they cannot be continued to non-perturbed plane Couette flow
either. Our results therefore suggest that it is not possible to
identify two-dimensional nonlinear solutions in plane Couette flow using
these two particular methods.

 The paper is organized as folllows. In Sec.~\ref{equations}, we formulate
 the physical problem, describe the underlying symmetries and present
 the numerical techniques  used to compute nonlinear branches of
 solutions. Sections~\ref{homotopy1} and \ref{homotopy2} describe
 our efforts to obtain steady two-dimensional nonlinear plane Couette
 solutions using either the Poiseuille-Couette homotopy with the same
 procedure as CE95 or the symmetry-preserving homotopy of MH05.
 A short discussion of the results (Sec.~\ref{discussion}) concludes the
 paper.

\section{Equations, symmetries\label{equations}, numerics}
\subsection{Physical model and background state}
In the following, an incompressible fluid with kinematic viscosity $\nu$
and constant density $\rho$ taken to be unity is considered.
The plane shear flows investigated here are independent of the
streamwise coordinate $x$ and are
bounded by two rigid walls in the shearwise $y$ direction, with
$-d<y<d$, where $d$ is chosen as length unit. 
The $x$ direction is taken to be periodic with spatial period $L_x=2\pi
d/\alpha$. A reference velocity $U_o$ is defined and a mean streamwise
pressure gradient $-2\nu\,U_o(1-\eta)/d^2$ and mean streamwise wall
velocities $U_B(y=\pm 1)=\pm \eta\,U_o$ are imposed in order to maintain
the following generic (non-dimensional) streamwise oriented background shear flow:
\begin{equation}
  \label{eq:bf}
U_B(y)=\left(1-\eta\right)\left(1-y^2\right)+\eta\,y~.   
\end{equation}
Here, $\eta$ is a nondimensional parameter that allows continuation from
plane Poiseuille flow ($\eta=0$) to plane Couette flow  ($\eta=1$). The
Reynolds number of the flow, based on $U_o$ and  half of the channel
width, reads  $\rey=(U_o d)/\nu$.
\subsection{Equations} 
The full non-dimensional velocity field, denoted by $\vec{U}$, is decomposed into
the original streamwise background flow $(U_B,0)$ and two-dimensional velocity
perturbations denoted by $\vec{u}=(u,v)$. The corresponding total pressure is
$P=P_B+p$, where $p$ is a pressure perturbation, and  the nondimensional
equations governing the evolution of
$\vec{U}$ and $P$ are the incompressible Navier-Stokes equations
\begin{equation}
\label{eq:NS}
\partial_t\,\vec{U}+\vec{U}\cdot\vec{\nabla}\vec{U}%+\ro\,\vec{e}_z\times\vec{u}
=-\grad{P}+\frac{1}{\rey}\Delta\vec{U}~,\quad\div{\vec{U}}=0.
\end{equation}
They are supplemented by periodic boundary conditions in $x$ 
and by no-slip boundary conditions in $y$ 
for $u$ and $v$:
\begin{equation}
\label{eq:bc}
u(x,-1,t )=u(x,1,t)=v(x,-1,t)=v(x,1,t)=0~,
%\displaystyle{v(x,-1)=v(x,1)=\diffy{v}(x,-1)=\diffy{v}(x,1)=0~.}
%\end{array}
\end{equation}
\noindent For numerical purposes, a Fourier decomposition 
is used in the $x$ direction and a scalar equation for the shearwise
velocity $v$ (or equivalently for a stream function $\psi$) governing
the evolution of the $k_x\neq 0$ modes is solved:
\begin{equation}
\label{eq:v}
\left(\difft{}-\f{1}{\rey}\Delta\right)\Delta
 v+\vec{P}_v\cdot\left(%
%\ro\vec{e}_z\times\vec{u}+
\vec{U}\cdot\grad{\vec{U}}\right)=0~,
\end{equation}
% \begin{equation}
% \label{eq:eta}
%    \left(\difft{}-\f{1}{\rey}\Delta\right)\eta+\vec{P}_\eta\cdot\left(%
% %\ro\vec{e}_z\times\vec{u}+
% \vec{u}\cdot\grad{\vec{u}}\right)=0~,
% \end{equation}
where $\vec{P}_v = -\vec{e}_y\cdot\curl{\curl{(\cdot)}}$.
%\\
%\begin{eqnarray}
%  \label{eq:PvPeta}
%  \vec{P}_\eta & = & \vec{e}_y\cdot\curl{(\cdot)}~.
%\end{eqnarray}
An auxiliary equation describing the evolution of the mean streamwise
velocity field must be solved simultaneously for completeness:
\begin{equation}
  \label{eq:meanflow}
 \rey \left(\difft{}\overline{u}+\partial_y\overline{uv}\right)=\ddiffyy{}\overline{u}~,
\end{equation}
where overbars denote $x$-averaged quantities.
\subsection{Symmetries}
Depending on the value of $\eta$, this set of equations has different
symmetries. For $\eta\neq 1$, the only symmetry is the translation symmetry
$SO(2)$
 \begin{equation}
   \label{eq:so2}
   SO(2): (x,y)\rightarrow (x+a,y)~,\quad(u,v,p)\rightarrow (u,v,p)
 \end{equation}
 with $a \in \left[0,2\pi/\alpha\right]$. 
For plane Couette flow ($\eta=1$), an extra (point) reflection symmetry
$Z^2$ is present,
 \begin{equation}
   \label{eq:Z2}
  Z^2: (x,y)\rightarrow (-x,-y)~,\quad (u,v,p)\rightarrow (-u,-v,p) ~.
 \end{equation}
In that case, the full flow has the more general $O(2)$ symmetry. %Note
%that the $Z^2$ symmetry, when present, can be used to reduce 
%computational costs.
\subsection{A symmetry preserving homotopy}
An extra symmetry-preserving artificial term proportional to $\eps
\vec{u}$, where $\eps$ is a small non-dimensional parameter, can be added to the
r.h.s. of the original Navier-Stokes equations~(\ref{eq:NS}) in order to
cause the least stable eigenvalue to become unstable and to perform the
symmetry-preserving homotopy mentioned in the introduction. This
technique is described in detail in MH05.
%  Once such nonlinear solutions
% have been obtained, we hope to be able to continue them down to
% $\epsilon=0$ by progressively reducing $\epsilon$. Unlike the
% Poiseuille-Couette homotopy, which consists of a variation of the $\eta$
% parameter in equation~(\ref{eq:bf}), the $\epsilon$ homotopy has the
% great advantage of  preserving all the original symmetries of the
% problem under consideration.
\subsection{Numerics\label{numerics}}
The exact two-dimensional nonlinear steady and travelling solutions of
Eqs.~(\ref{eq:bc})-(\ref{eq:meanflow}) presented in
the next sections have been obtained thanks to a nonlinear continuation code
\citep{rincon07} similar to that of \citet{waleffe03} and \citet{wedin04}.
Chebyshev collocation on a Gauss-Lobatto grid  in the shearwise
direction and a Fourier representation  in the $x$ 
direction are used. Clamped boundary conditions are implemented using
differentiation matrices computed with the DMSuite
package\citep{weidemann00}. Nonlinear terms are computed in physical
space and steady nonlinear equations are solved in
spectral-physical-spectral space using Newton iteration. A direct LAPACK
solver is called at each Newton iteration and dealiasing is performed
in the $x$-direction.  Pseudo-arclength continuation \citep{boyd} 
is implemented to follow branches of solutions.
The code has been tested quantitatively by solving the classical
problem of subcritical continuation of two-dimensional travelling waves
in no-slip plane Poiseuille flow \citep{zahn74,herbert76}. The full
three-dimensional version (not used here) also quantitatively reproduces
the continuation diagrams of three-dimensional nonlinear solutions
obtained by \citet{waleffe03} in no-slip, non-rotating plane Couette
flow\citep{rincon07}. 

% Results obtained with these programs were
% then compared with the results of MH05 obtained with a code using an implicit
% streamfunction formulation of the Orr-Sommerfeld equations and the
% software package AUTO \citep{auto97} to compute nonlinear branches of
% solutions \citep{mehta04}.

% Steady solutions of eq. ??? + Phase condition + travelling waves

\section{Poiseuille-Couette homotopy\label{homotopy1}}
The Poiseuille-Couette homotopy was first considered in order to compute
the two-dimensional nonlinear solutions reported by CE95 in plane
Couette flow. No particular assumption regarding
the symmetries of the solutions was made in order to perform these
computations and the numerical method employed
here copes with solutions with either zero or non-zero phase
speed (the transformation $\difft{}=-c\,\partial_x$, where $c$ is the
phase speed, is used in the latter case). 
All our attempts proved unsuccessful. We simply followed their procedure
and performed continuation with respect to $\eta$,
starting from two-dimensional nonlinear travelling wave solutions in
Poiseuille flow ($\eta=0$) at $\rey=15993$, which corresponds to the
first case considered by CE95.
Continuation curves for mixed Poiseuille-Couette flow are shown in
Fig.~\ref{fig1}. For the maximum affordable resolution and $\alpha=1$, a
turning point is found at $\eta\simeq
0.23$, therefore it proves impossible to isolate solutions for plane Couette
flow. Since CE95 found Couette solutions for small $\alpha$ only (albeit
for smaller $\rey$),  we tried to perform further continuation with respect
to $\alpha$ for various values of $\eta$ and different resolutions
including the maximum resolution  used by CE95 (corresponding
to $N_x=52$ in our code). The results, presented in Fig.~\ref{fig2},
show that it is not possible to identify plane Couette flow solutions
for small values of $\alpha$ either. The nose of the surface of solutions in
the $\alpha-\eta-$ wall shear rate space is located at $\eta\simeq 0.5$
and $\alpha\simeq 0.2$ for $\rey=15993$, even for rather large
streamwise resolution. This result is similar to
Fig.~1 of CE95, who presented low streamwise resolution solutions down to
$\eta=0.45$ for $\rey=15993$. As already acknowledged in CE95, it is
clear from the figure that there are convergence issues at low $\alpha$
for such a high value of $\rey$, even with a resolution such as
$(N_x,N_y)=(52,48)$. We notably encountered difficulties performing the 
low-$\alpha$ continuation at $\eta=0.45$.  The same procedure was 
consequently attempted at the less resolution-demanding $\rey=7003$ 
at which CE95 found their Couette flow solutions. 
The corresponding continuation curves for mixed Poiseuille-Couette flow
are also shown in Fig.~\ref{fig1}. This time, a turning point is found at
$\eta\simeq 0.14$ for $\alpha=1$. Further
continuations with respect to $\alpha$ for various values of $\eta$ and
the same maximum resolution as the one used by CE95 are represented in
Fig.~\ref{fig3} and show that it is not possible to identify plane
Couette flow solutions for small values of
$\alpha$ either. The nose of the surface of solutions in the
$\alpha-\eta-$wall shear rate space is clearly located at $\eta\simeq
0.14$ and $\alpha\simeq 1$ for $\rey=7003$.%  Low-resolution ($N_x=6$),
                                           %  dealiased results are also
% depicted in Fig.~\ref{fig3} and  Fig.~\ref{fig4} and exhibit similar
% behaviour. 
 The same procedure was finally attempted for even smaller
values of $\rey$ and led to similar negative results. 

\begin{figure}[t]
\resizebox{\hsize}{!}{%
\includegraphics[width=9.5cm]{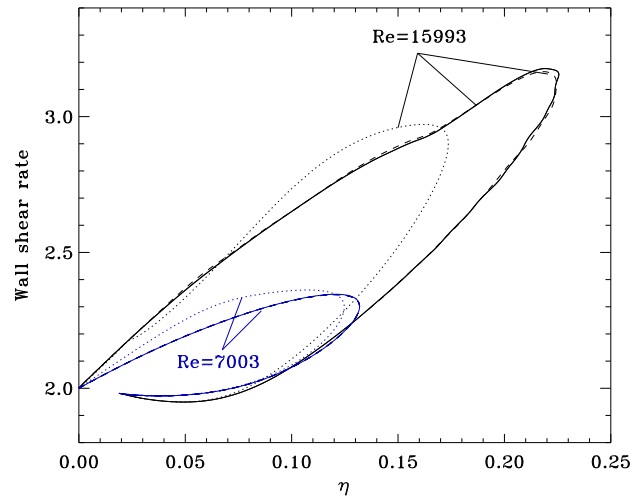}}
\caption{Continuation of two-dimensional solutions with $\alpha=1$ in
         mixed Poiseuille-Couette flow~(\ref{eq:bf})  as a function of 
        $\eta$, for $\rey=15993$ (black curves with turning points at
        $\eta\simeq 0.23$ and  $\rey=7003$ ({\color{blue}blue} online,
        with turning points at $\eta\simeq 0.13$). Full line:
        $(N_x,N_y)=(52,48)$, dashed line (almost undistinguishable from
        full line):  
        $(N_x,N_y)=(32,48)$, dotted line: $(N_x,N_y)=(6,48)$. The
        non-dimensional wall shear rate $|\partial
     \overline{U}/\partial y|(y=\pm 1)$ has been used as a measure of
     nonlinearity (for the laminar flow, $|\partial \overline{U}/\partial
     y|(\pm 1)=1$). The "gap" at $\eta\simeq 0$ results from the existence
     of a turning point close to $\eta\simeq 0.02$.}
\label{fig1}
\end{figure}

The main trend  is therefore that it seems  possible to reach solutions with lower $\eta$
with increasing $\rey$, but it is clear from our results that no
$\eta=1$ solution can be found for the range of $\rey$ investigated by
CE95. Since only very little information is given on their precise procedure to obtain
two-dimensional solutions, it is actually very hard to figure
out where the problem could come from in their results.  We originally thought
that aliasing errors may be present in their calculations, but since
the numerical method used by CE95 is fully spectral, their
results should not in principle  suffer from this problem, at least in
the streamwise direction. One possibility would be that their error
tolerance for the Newton  algorithm was not small enough. In the present
calculations, convergence was considered to be achieved for an error
smaller than $10^{-7}$ in energy and a precision of $10^{-9}$ was
reached in many cases.
To conclude this section, it is worth noting that the
\citet{milinazzo85} and CE95 homotopy~(\ref{eq:bf}) is only a particular
choice of Poiseuille-Couette homotopy, and that more general
transformations from one flow to the other, \textit{e.~g.}
\begin{equation}
  \label{eq:bf2}
U_B(y)=\left(1-f(\eta)\right)\left(1-y^2\right)+g(\eta)\,y~,
\end{equation}
where $f$ and $g$ satisfy the right end conditions, may lead to
different results. This kind of behaviour has notably been reported
in Taylor-Couette flow by \citet{ruck04}, who discovered that a
connection between solutions for two different sets of boundary
conditions was only possible  in their problem  for some special
homotopy choices.

\begin{figure}
\resizebox{\hsize}{!}{%
\includegraphics[width=9.5cm]{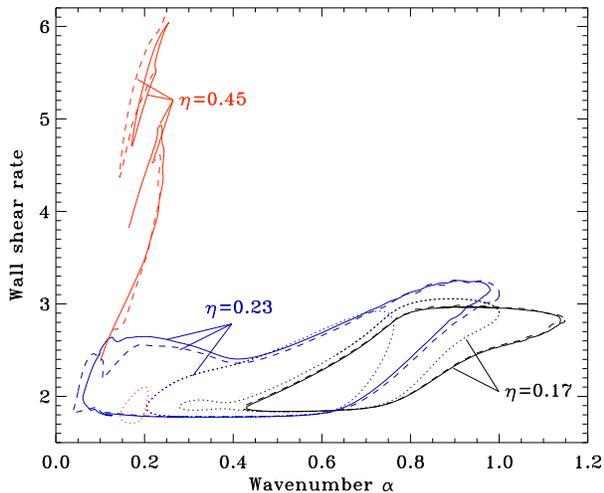}}
\caption{Continuation of two-dimensional solutions ($\rey=15993$) in mixed
        Poiseuille-Couette flow~(\ref{eq:bf})  as a function of $\alpha$ for
        $\eta=0.17$ (black curves with the largest $\alpha$ values),
        $\eta=0.23$ ({\color{blue}blue} online, curves centered on $\alpha\simeq 0.5$),
        $\eta=0.45$ ({\color{red}red} online, narrow curves at low
        $\alpha$. Full line: $(N_x,N_y)=(52,48)$,   dashed line:
        $(N_x,N_y)=(32,48)$,  dotted line:$(N_x,N_y)=(6,48)$. Solutions
        for $N_x=6$ do not extend to $\eta=0.45$ and the corresponding red
        dotted curve at $\alpha\simeq 0.2$ is for $\eta=0.4$.  Note the
        convergence problems at low $\alpha$ for $(N_x,N_y)=(32,48)$ and
        $(N_x,N_y)=(52,48)$.}
\label{fig2}
\end{figure}

\begin{figure}
\resizebox{\hsize}{!}{%
\includegraphics[width=9.5cm]{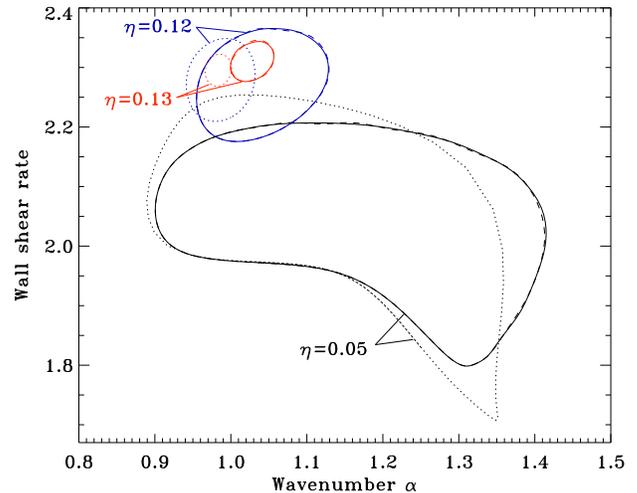}}
\caption{Continuation of two-dimensional solutions ($\rey=7003$) in mixed
        Poiseuille-Couette flow~(\ref{eq:bf})  as a function of $\alpha$ for
        $\eta=0.05$ (black, family of large curves), $\eta=0.12$ ({\color{blue}blue}
        online, family of medium size curves), $\eta=0.132$
        ({\color{red}red} online, family of small curves corresponding
        to the ``nose'' of the surface of
        solutions). Full line: $(N_x,N_y)=(52,48)$, 
        dashed line (undistinguishable from full line curves): $(N_x,N_y)=(32,48)$, 
        dotted line:$(N_x,N_y)=(6,48)$.
        The results for $(N_x,N_y)=(52,48)$ presented here are also
        undistinguishable from those (not shown here) obtained for
        the same resolution $(N_x,N_y)=(52,28)$ as that used by CE95.}
\label{fig3}
\end{figure}

\section{Symmetry-preserving homotopy\label{homotopy2}}
\begin{figure}
\resizebox{\hsize}{!}{%
\includegraphics[width=9.5cm]{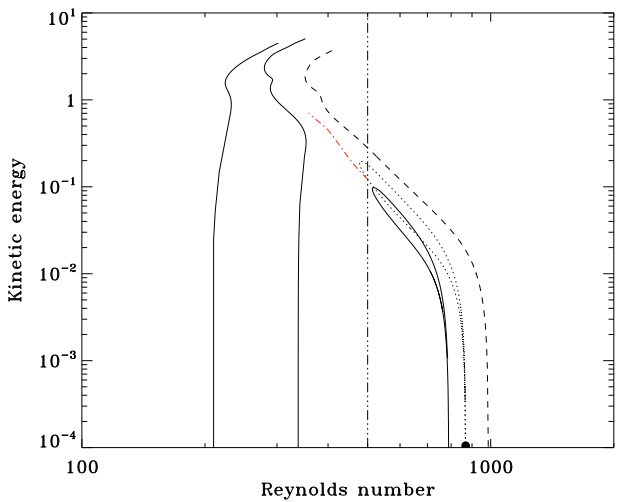}}
\caption{Bifurcations and nonlinear continuation of 
  two-dimensional solutions of the  $\eps$-perturbed plane Couette flow of MH05 
  with $\alpha=0.2$,  $\eps=0.05$  (full line), $\eps=0.04$ (dashed line) and 
  $\eps=0.04547$ (dotted and dash-dotted lines). $(N_x,N_y)=(32,48)$ has been 
  used (the results of some of these computations done at $(N_x,N_y)=(52,48)$ are
  undistinguishable from those at $N_x=32$).
  All solutions are steady except for the dash-dotted line branch at
  $\eps=0.04547$ (dash-dotted, {\color{red}red} online),
  for which the phase speed reaches 0 when it meets the dotted steady
  branches. The vertical dash-triple dotted line corresponds to the path
  followed in the continuation presented in Fig.~\ref{fig5}.
  Continuation was stopped at large energy (strong nonlinearity) due to
  insufficient numerical resolution.}
\label{fig4} 
\end{figure}

Motivated by the apparent similarities between the steady solutions of CE95 and
those that MH05 obtained for small values of their perturbation
parameter $\eps$, we then considered the
problem of finding steady two-dimensional
nonlinear solutions in plane Couette flow ($\eta=1$) using the homotopy proposed by
MH05 and parameters similar to those used in their calculations.
An $\eps$ symmetry-preserving identity perturbation was included
in the equations for that purpose. As mentioned in Sec.~\ref{equations},
this causes the least stable eigenvalue of the Couette flow to become
unstable. For $\rey>375$ and $\alpha=0.2$, the bifurcation is of the
Hopf type, while for $\rey<375$ and the same $\alpha$, the bifurcation is steady.
In the latter case, for an identity perturbation with $\eps>\eps_0$,
where $\eps_0\simeq 0.0456$, the eigenvalue crosses the imaginary axis
for two values of $\rey$ (see MH05), leading to two different branches of steady
nonlinear solutions (Fig.~\ref{fig4}). MH05 showed that steady branches with $\eps<\eps_0$
cannot be computed directly  because they reduce to a single point at
$\eps=\eps_0$ and $\rey\simeq 270$ (see their Fig.~11). Nevertheless,
as noted in their paper, there is a connection  between low $\rey$
branches with $\eps>\eps_0$ and high $\rey$ branches, including some 
with $\eps<\eps_0$. For instance, a solution at $\eps=0.04$ (dashed
line) can be obtained by reducing $\eps$, starting from a high energy 
solution at $\eps=0.05$ and $\rey\simeq 360$. The associated $\eps=0.04$
steady branch can then be continued to large values of $\rey$ and
transformed into high $\rey$ branches with different values of
$\eps$. 

\begin{figure}
\resizebox{\hsize}{!}{%
\includegraphics[width=9.5cm]{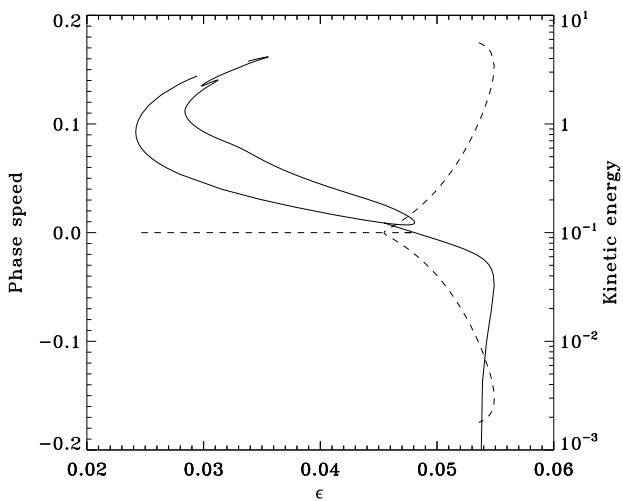}}
\caption{Nonlinear continuation of two-dimensional solutions of the $\eps$-perturbed 
  plane Couette flow of MH05 with respect to $\eps$, for $\alpha=0.2$, $\rey=500$ and
  $(N_x,N_y)=(32,48)$. The dashed line 
  corresponds to the phase speed of the solutions and the full line 
  to their kinetic energy. The starting point of the continuation
  corresponds to the Hopf bifurcation of the perturbed Couette flow,
  which occurs at $\eps=0.05358$ for $\rey=500$. Continuation was 
  stopped at large energy (strong nonlinearity) due to 
  insufficient numerical resolution.}
\label{fig5} 
\end{figure}

 \begin{figure}[h]
% \vspace{5cm}
\resizebox{\hsize}{!}{%
\includegraphics[width=9.5cm]{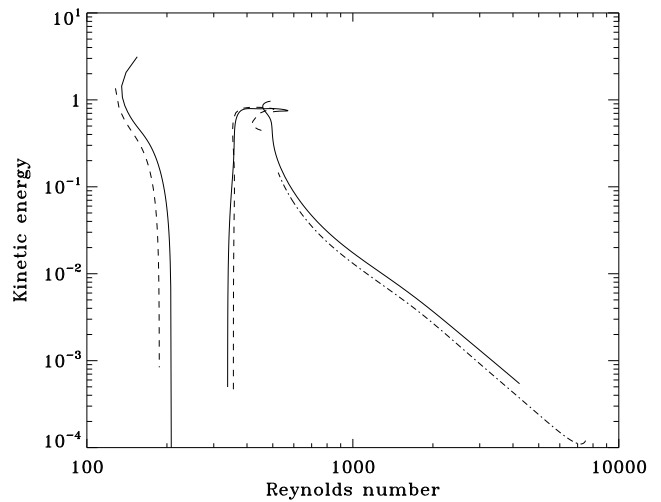}}
 \caption{Bifurcations and nonlinear continuation of steady
   two-dimensional $\alpha=0.2$ solutions of the $\eps$-perturbed plane Couette flow
   without the artifact mean flow perturbation for
   $\eps=0.05$   (full line), $\eps=0.05358$
   (dashed line) and $\eps=0.04547$ (dash-dotted line) when the $\eps$
   perturbation term is added to the  $k_x\neq 0$ part of the equations
   only. $(N_x,N_y)=(32,48)$. Continuation was stopped at large energy 
  (strong nonlinearity) due to insufficient numerical resolution.}
 \label{fig6}
 \end{figure}

In Fig.~\ref{fig4}, a travelling branch at $\eps=0.04547$  is also
plotted. This branch connects to a steady high $\rey$ branch with
the same $\eps$ at $\rey=500$. The presence of such a travelling branch,
which was discovered by performing continuation with respect to $\eps$ in a
range of $\rey$ where steady branches of solutions also exist, confirms
the findings of MH05 that travelling solutions bifurcating at $\rey>375$
can turn into steady solutions by changing $\eps$ at fixed $\rey$.
Fig.~\ref{fig5} shows a typical continuation with respect to $\eps$
that clarifies this connection. Continuation in this figure is done
along the line $\rey=500$ represented in Fig.~\ref{fig4}.
One finds that there are in fact two travelling branches of solutions
with the same kinetic energy and opposite phase speed. These solutions
merge at $\eps=0.04547$ and transform into a family of nonlinear steady
solutions which corresponds to a cut at fixed $\rey$ of the surface of
solutions in the $\rey-\eps-\mbox{amplitude}$ space. We conjecture that
this behaviour is a reminiscence in the nonlinear regime
of the collision between the first two  real eigenvalues of the linear
problem and their subsequent splitting into a  complex conjugate pair
(see MH05 for full details regarding the linear problem).
  Note finally that Fig.~\ref{fig5} is in qualitative agreement
  \footnote{In the course of this study, P.~G.~Mehta (private
  communication) pointed out to us that a lack of quantitative agreement
  is not surprising here due to the limited resolution capacities of
  the code used by MH05. Note that the code used in the present
  paper gives robust results even for very large systems of equations 
  (including three-dimensional problems)  and that it
  quantitatively reproduces the steady results of MH05 (compare
  Fig.~\ref{fig4} here and their Fig.~12).} with Fig.~6 of MH05.

The behaviour of  steady solutions at large $\rey$
suggests that solutions with smaller and smaller values of $\eps$ can be found 
provided that $\rey$ is increased sufficiently. However, a careful
examination of Fig.~\ref{fig4} reveals the presence of intriguing ``double''
bifurcations at large $\rey$ (see for instance the closed loop in dotted line 
emanating from the filled black circle at $\rey=868.2 $ and $\eps=0.04547$)
that are not predicted by standard linear stability
analysis for infinitesimal velocity perturbations with a single
wavenumber $\alpha\neq 0$  (this also occurs in Fig.~12 of MH05).
This  behaviour can be explained as follows. Consider the
equation for the perturbed mean flow 
\begin{equation}
  \label{eq:meanflow2}
\rey\left(\partial_t \overline{u}+\partial_y\overline{uv}\right)
= \partial^2_y{\overline{u}}+\eps\rey\overline{u},
\end{equation}
supplemented by the set of boundary conditions~(\ref{eq:bc}) on $u$.
Eigenmodes of the linearized version of this equation take on the form
\begin{equation}
  \label{eq:meanflow3}
  \overline{u}=A\sin(n\pi y),
\end{equation}
and their growth rate $\gamma$ is given by 
\begin{equation}
\label{eq:growthrate}
\gamma=\eps-(n\pi)^2/\rey~. 
\end{equation}
These linear modes have no self-interactions, therefore they are also
exact steady solutions of the fully nonlinear equations for any amplitude $A$ when
$\eps\rey=(n\pi)^2$. For $n=2$, we precisely obtain the bifurcation points
observed in Fig.~\ref{fig4} at $\rey=868.2$ for $\eps=0.04547$ and at
$\rey=789.6$ for $\eps=0.05$, which hints that these mean flow modes 
participate in the dynamics of high $\rey$ solutions described
earlier. Note that the original idea of MH05 was to introduce an $\eps$ perturbation
in order to cause the least stable  eigenmode of Plane Couette flow with
$k_x=\alpha$ to bifurcate, not to influence the whole dynamics of the
flow including that of the $k_x=0$ Fourier mode.

 \begin{figure}[t!]
% \vspace{5cm}
\resizebox{\hsize}{!}{%
\includegraphics[width=9.5cm]{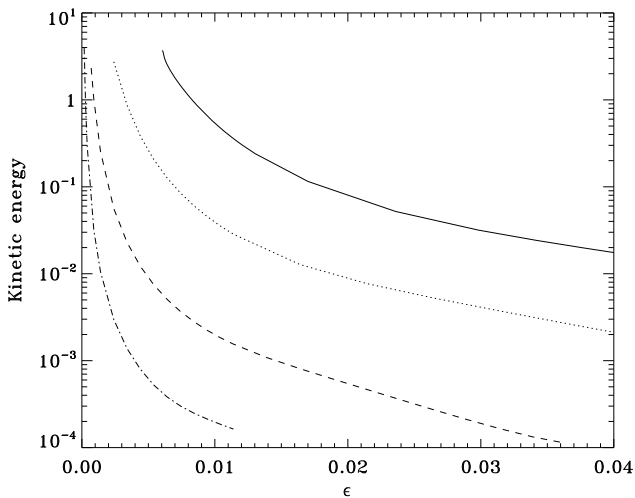}}
 \caption{Nonlinear continuation of steady $\alpha=0.2$ two-dimensional solutions
   of the $\eps$-perturbed plane Couette flow without the artifact mean flow perturbation 
   as a function of $\eps$, for $\rey=1000$  (full line), $\rey=2500$ (dotted
   line), $\rey=9\,000$ (dashed line) and  $\rey=30\,000$ (dash-dotted
   line). $(N_x,N_y)=(32,48)$.}
 \label{fig7}
 \end{figure}

It is therefore natural to ask whether or not high
$\rey$, low $\eps$ branches survive when the artificial mean flow
eigenmode is removed, and if solutions with $\eps=0$ can be found
subsequently. Thanks to the way the problem has
been set-up numerically in this study, it is straightforward to get rid of this
model artefact by adding the $\eps$ perturbation in Eq.~(\ref{eq:v}) but
not in Eq.~(\ref{eq:meanflow}).  Fig.~\ref{fig6} shows that branches of
solutions at small $\rey$ computed with the corrected model look similar
to those in Fig.~\ref{fig4}, which were obtained with the original
MH05 model. The main difference is for the lowest $\rey$ branch represented in
these figures, which is supercritical in the MH05 model whereas it is subcritical
when the mean flow perturbation is removed.
A connection with high $\rey$ branches is also found in
the corrected model, but in that case the branches do not 
bifurcate at some $\eps$-dependent critical $\rey$, even though they
have a rather low kinetic energy at large $\rey$. In order to
discover nonlinear plane Couette flow solutions, we finally attempted to
continue solutions obtained at a given $\rey$  with respect to $\eps$. This
continuation is presented in Fig.~\ref{fig7} for different values of
$\rey$. The amplitude of the solutions
is seen to asymptote to infinity in the limit of vanishing $\eps$
for all selected values of $\rey$. The natural conclusion  is that no
two-dimensional nonlinear solutions can be found for plane Couette flow
using this homotopy technique.

\section{Conclusions\label{discussion}}
The results presented in this study tend to show that one cannot
identify steady nonlinear two-dimensional solutions in plane
Couette flow using either the Poiseuille-Couette homotopy of CE95 or
the $\eps$-perturbation homotopy suggested by MH05. We have been unable
to recover  the solutions reported by CE95 and have pointed out
that they may be spurious. It is therefore also likely that the results
of \citet{ehren97} suffer from the same problem. Other more general homotopies like
Eq.~(\ref{eq:bf2}) may however be helpful to discover a connection
between  Poiseuille and Couette flow. The steady solutions
obtained by MH05 at large $\rey$ and small $\eps$ values were shown to
be influenced by an artifact of the perturbed equations, however using
a revised version of their model did not help discovering new solutions 
for plane Couette flow because the corrected solutions found
at small $\eps$ asymptote to infinite amplitude in the  $\eps=0$ limit. 
This leaves the issue of the existence of two-dimensional nonlinear
solutions in plane Couette flow unsettled. 
Another interesting avenue of investigation
would be to consider the fate of the inviscid solutions discovered by
\citet{Kida81} in a uniform unbounded shear flow  when a small
amount of viscosity is added. The relevance of this type of
solutions to wall-bounded flows is however quite unclear. 

To conclude, it is worth pointing out that discovering nonlinear
two-dimensional solutions in plane {Couette} flow is also relevant
to the problem of the nonlinear stability of linearly stable swirling
flows. Indeed, plane Couette flow with constant (spanwise) rotation can
be used as a local model of differentially rotating flows. Since two-dimensional
incompressible dynamics in plane Couette flow is not affected by the
presence of a Coriolis force associated with a  rotation vector
perpendicular to the  $(x,y)$ plane,  possible two-dimensional
nonlinear solutions  in non-rotating plane Couette flow would also be
solutions in plane Couette flow with constant
spanwise rotation for any rotation regime, including centrifugally stable
anticyclonic regimes for which no purely hydrodynamic solutions
are currently known at all\citep{rincon07}.

\section*{Acknowledgments}
%\begin{acknowledgments}
The author acknowledges several helpful discussions with P.~G. Mehta,
G.~I. Ogilvie and A.~M. Rucklidge. This work has been supported by the
Leverhulme Trust and the Newton Trust.
%\end{acknowledgments}

\bibliographystyle{unsrtnat}
\bibliography{couette_final}

\begin{thebibliography}{17}
\providecommand{\natexlab}[1]{#1}
\providecommand{\url}[1]{\texttt{#1}}
\expandafter\ifx\csname urlstyle\endcsname\relax
  \providecommand{\doi}[1]{doi: #1}\else
  \providecommand{\doi}{doi: \begingroup \urlstyle{rm}\Url}\fi

\bibitem[{Bech} et~al.(1995){Bech}, {Tillmark}, {Alfredsson}, and
  {Andersson}]{bech95}
K.~H. {Bech}, N.~{Tillmark}, P.~H. {Alfredsson}, and H.~I. {Andersson}.
\newblock An investigation of turbulent plane couette flow at low reynolds
  numbers.
\newblock \emph{J. Fluid Mech.}, 286:\penalty0 291--325, 1995.

\bibitem[{Nagata}(1990)]{nagata90}
M.~{Nagata}.
\newblock Three-dimensional finite-amplitude solutions in plane {C}ouette flow:
  bifurcation from infinity.
\newblock \emph{J. Fluid Mech.}, 217:\penalty0 519--527, 1990.

\bibitem[{Clever} and {Busse}(1992)]{clever92}
R.~M. {Clever} and F.~H. {Busse}.
\newblock Three-dimensional convection in a horizontal fluid layer subjected to
  a constant shear.
\newblock \emph{J. Fluid Mech.}, 234:\penalty0 511--527, 1992.

\bibitem[{Waleffe}(2001)]{waleffe01}
F.~{Waleffe}.
\newblock {Exact coherent structures in channel flow}.
\newblock \emph{J. Fluid Mech.}, 435:\penalty0 93--102, 2001.

\bibitem[{Waleffe}(2003)]{waleffe03}
F.~{Waleffe}.
\newblock {Homotopy of exact coherent structures in plane shear flows}.
\newblock \emph{Phys. Fluids}, 15:\penalty0 1517--1534, 2003.

\bibitem[{Milinazzo} and {Saffman}(1985)]{milinazzo85}
F.~A. {Milinazzo} and P.~G. {Saffman}.
\newblock Finite-amplitude steady waves in plane viscous shear flows.
\newblock \emph{J. Fluid Mech.}, 160:\penalty0 280, 1985.

\bibitem[{Zahn} et~al.(1974){Zahn}, {Toomre}, {Spiegel}, and {Gough}]{zahn74}
J.~P. {Zahn}, J.~{Toomre}, E.~A. {Spiegel}, and D.~O. {Gough}.
\newblock Nonlinear cellular motions in {P}oiseuille flow.
\newblock \emph{J. Fluid Mech.}, 64:\penalty0 319--345, 1974.

\bibitem[{Herbert}(1976)]{herbert76}
T.~{Herbert}.
\newblock {Periodic secondary motions in a plane channel}.
\newblock In A.~I. {van de Vooren} and P.~J. {Zandbergen}, editors,
  \emph{{P}roc. {I}nt. {C}onf. {N}umer. {M}ethods {F}luid {D}yn.}, pages
  235--240. {B}erlin: {S}pringer-{V}erlag, 1976.

\bibitem[{Cherhabili} and {Ehrenstein}(1995)]{ehren95}
A.~{Cherhabili} and U.~{Ehrenstein}.
\newblock Spatially localized two-dimensional finite-amplitude states in plane
  {Couette} flow.
\newblock \emph{Eur. J. Mech. B/Fluids}, 14:\penalty0 667, 1995.

\bibitem[{Cherhabili} and {Ehrenstein}(1997)]{ehren97}
A.~{Cherhabili} and U.~{Ehrenstein}.
\newblock {Finite-amplitude equilibrium states in plane {Couette} flow}.
\newblock \emph{J. Fluid Mech.}, 342:\penalty0 159--177, 1997.

\bibitem[{Mehta} and {Healey}(2005)]{mehta05}
P.~G. {Mehta} and T.~J. {Healey}.
\newblock {On steady solutions of symmetry-preserving perturbations of the
  two-dimensional {Couette} flow problem}.
\newblock \emph{Phys. Fluids}, 17:\penalty0 4108, 2005.

\bibitem[{Rincon} et~al.(2007){Rincon}, {Ogilvie}, and {Cossu}]{rincon07}
F.~{Rincon}, G.~I. {Ogilvie}, and C.~{Cossu}.
\newblock On self-sustaining processes in {Rayleigh}-stable rotating plane
  {Couette} flows and subcritical transition to turbulence in accretion disks.
\newblock \emph{Astron. \& Astrophys.}, 463:\penalty0 817--832, 2007.

\bibitem[{Wedin} and {Kerswell}(2004)]{wedin04}
H.~{Wedin} and R.~R. {Kerswell}.
\newblock {Exact coherent structures in pipe flow: travelling wave solutions}.
\newblock \emph{J. Fluid Mech.}, 508:\penalty0 333--371, 2004.

\bibitem[{Weidemann} and {Reddy}(2000)]{weidemann00}
J.~A.~C. {Weidemann} and S.~C. {Reddy}.
\newblock {A Matlab differentiation suite}.
\newblock \emph{{ACM Transactions on Mathematical Software}}, 26\penalty0
  (4):\penalty0 465--519, 2000.

\bibitem[{Boyd}(2001)]{boyd}
J.~P. {Boyd}.
\newblock \emph{Chebyshev and Fourier Spectral Methods}.
\newblock Dover, Mineola, New York, 2001.
\newblock Second Edition.

\bibitem[{Rucklidge} and {Champneys}(2004)]{ruck04}
A.~M. {Rucklidge} and A.~R. {Champneys}.
\newblock Boundary effects and the onset of {Taylor} vortices.
\newblock \emph{Physica D}, 191:\penalty0 282--296, 2004.

\bibitem[Kida(1981)]{Kida81}
S.~Kida.
\newblock Motion of an elliptic vortex in a uniform shear flow.
\newblock \emph{J. Phys. Soc. Japan}, 50:\penalty0 3517, 1981.

\end{thebibliography}

\end{document}